\documentclass[aps,10pt,floatfix,twocolumn,showpacs,superscriptaddress]{revtex4-1} 
\usepackage{color,soul}
\usepackage{graphicx}
\usepackage{siunitx}
\usepackage[breaklinks=true,colorlinks=true,linkcolor=blue,urlcolor=blue,citecolor=blue]{hyperref}

\usepackage{dblfloatfix}
\usepackage{amsmath, amsthm, amssymb}
\usepackage{amsthm}
\usepackage{multirow}
\usepackage[normalem]{ulem}
\usepackage{soul}
\usepackage{makecell}
\usepackage[table]{xcolor}
\definecolor{Gray}{gray}{0.85}
\definecolor{LightCyan}{rgb}{0.88, 1, 1}
\definecolor{Apricot}{rgb}{0.98, 0.81, 0.69}
\usepackage{threeparttable}
\newcommand{\be}{\begin{equation}}
\newcommand{\ee}{\end{equation}}
\newcommand{\bea}{\begin{eqnarray}}
\newcommand{\eea}{\end{eqnarray}}

\usepackage[rightcaption,]{sidecap}
\sidecaptionvpos{figure}{c}
\usepackage[utf8]{inputenc}
\usepackage{amsmath}
\usepackage{amssymb}
\usepackage{graphicx}
\usepackage{hyperref}
\usepackage[export]{adjustbox}
\usepackage[update]{epstopdf}
\usepackage{multirow}
\usepackage{placeins}

\begin{document}

\title{Solvent-cosolvent attraction is sufficient to induce polymer collapse in good solvent mixtures}

\author{Hitesh Garg}
\email{hiteshgarg@imsc.res.in}
\affiliation{The Institute of Mathematical Sciences, C.I.T. Campus,
Taramani, Chennai 600113, India}
\affiliation{Homi Bhabha National Institute, Training School Complex, Anushakti Nagar, Mumbai, 400094, India}

\author{Divya Nayar}
\email{divyanayar@mse.iitd.ac.in}
\affiliation{Department of Materials Science and Engineering, Indian Institute of Technology, Delhi, India}

\author{Satyavani Vemparala}
\email{vani@imsc.res.in}
\affiliation{The Institute of Mathematical Sciences, C.I.T. Campus,
Taramani, Chennai 600113, India}
\affiliation{Homi Bhabha National Institute, Training School Complex, Anushakti Nagar, Mumbai, 400094, India}
\date{\today}
\begin{abstract}
Cononsolvency occurs when two miscible, competing good solvents for a polymer are mixed, resulting in a loss of solubility. In this study, we demonstrate through simulations, supported by theory, that cononsolvency can be driven solely by solvent-cosolvent attraction ($\epsilon_{sc}$). The primary mechanism underlying this behavior is the emergent depletion effect, which is amplified by solvent-cosolvent interactions. The polymer reaches a compact state when the solvent and cosolvent fractions are equal ($x_s = x_c = 0.5$), a finding that aligns with predictions from Flory-Huggins theory and the random phase approximation. We show that this cononsolvency behavior is observed for different cosolvent sizes, provided the cosolvent density remains below the depletion threshold and the sizes of solvent and cosolvent particles are not smaller than the monomer size. Additionally, we investigate the role of temperature and find that cononsolvency weakens as temperature increases, due to a reduction in the depletion effect. Finally, we show that when preferential cosolvent attraction is introduced in this simple model, it leads to cononsolvency driven by bridging interactions, occurring at lower cosolvent fractions ($x_c < 0.5$).
\end{abstract}

\maketitle
\section{Introduction}
The coil-to-globule transition in polymers can occur through various pathways, depending on the specific conditions and interactions at play. Common pathways include changes in solvent quality, the application of external pressure, and the addition of salts or other cosolutes, all of which can trigger conformational changes in polymers. Solvent quality is a particularly crucial factor, as modifications in the nature of the solvent can drive the coil-to-globule transition. In poor solvents, where polymer-solvent interactions are weaker, polymers tend to collapse, a transition often induced by increasing the fraction of non-solvents or altering solvent properties~\cite{flory1953principles, de1979scaling, de1990polymers}. Another pathway involves applying external pressure to reduce the available volume for the polymer chain, leading to collapse. Although less common than temperature-induced transitions, pressure-induced transitions are studied under specialized conditions~\cite{heremans1982high,canchi2013cosolvent}.

In addition, the presence of salts or cosolutes can influence the
coil-to-globule transition by modifying the ionic strength of the solution or
introducing specific interactions between polymer and solute particles. This is
especially relevant in the study of biopolymers like proteins and DNA, which
can undergo conformational changes in response to environmental factors such as
pH, ionic strength, or the presence of ligands and
cofactors~\cite{sherman2006coil, frerix2006exploitation, heyda2017guanidinium}. In biopolymers,
hydrophobic collapse is a crucial mechanism, where hydrophobic residues
aggregate to minimize contact with water, leading to the formation of compact,
globular structures, such as in protein
folding~\cite{clark2020water,zhou2004hydrophobic}. Additionally, charged
polymers can experience coil-to-globule transitions in response to changes in
electrostatic interactions, driven by alterations in the solution's ionic
strength and additional salt
concentration~\cite{wang2024cation,ashbaugh2008natively,sherman2006coil}. An
interesting environment for studying these transitions is the interior of
living cells, which are highly crowded with macromolecules and small cosolutes.
In such crowded conditions, depletion interactions can play a significant role
in inducing coil-to-globule transitions. Depletion occurs when smaller
particles or colloids act as depletants, leading to the collapse of polymer
chains into a more compact, globular
conformation~\cite{asakura1958interaction,asakura1954interaction,vrij1976polymers}.
This phenomenon, driven by the excluded volume effect, has practical
applications in colloid and polymer science, including the food industry.
Such a phenomenon has also be used to explain the biomolecular collapse-unfolding
transitions in biomolecules inside the crowded living cell.
Specific interactions between polymers and solvent/crowder particles, can also
lead to unexpected coil-to-globule transitions, particularly when attractive
crowders are involved, as demonstrated in several recent
studies~\cite{jiao2010attractive, kim2013crowding,
antypov2008computer,heyda2013rationalizing, rodriguez2015mechanism,
sagle2009investigating,huang2021chain, ryu2021bridging, brackley2020polymer,
brackley2013nonspecific,barbieri2012complexity, garg2023conformational,nayar2020small}.

In our recent work~\cite{garg2023conformational}, we showed that neutral
polymers with very weak self-interactions, when exposed to weakly attractive
crowders, undergo a distinct coil-to-globule transition as the crowder-crowder
attraction increases, depending on crowder density. Building on this
observation, we investigate the phenomenon of cononsolvency driven by
solvent-cosolvent attraction. Cononsolvency is a fascinating
and complex phenomenon, particularly relevant in mixed solvent
systems~\cite{wolf1978measured,magda1988dimensions,bischofberger2014co,bharadwaj2022cononsolvency,mukherji2014polymer,mukherji2015co}.
It occurs when a polymer, soluble in both components of a binary solvent
mixture, becomes insoluble when the solvents are mixed in specific proportions.
This counterintuitive behavior contrasts with the expected intermediate
solubility between the two pure solvents. The underlying mechanism of
cononsolvency is driven by intricate interactions among the polymer, solvent,
and cosolvent particles, influenced by factors such as the polymer's
hydrophilicity or hydrophobicity, solvent polarity, and thermodynamic
conditions like temperature and pressure. Several theories have been proposed
to explain cononsolvency, with a central focus on the balance between
polymer-solvent and solvent-(co)solvent interactions. In some cases, the addition
of a cosolvent disrupts favorable polymer-solvent interactions, resulting in
decreased solubility. Another significant factor is the change in solvent
quality as the composition varies; a cosolvent may alter the solvation
environment around the polymer, potentially leading to phase separation.
Moreover, entropic contributions from polymer-solvent interactions, along with
associated free energy changes in the solvent mixture, play pivotal roles in
driving the cononsolvency effect. This phenomenon has been extensively
investigated in systems such as water-alcohol mixtures with polymers like
poly(N-isopropylacrylamide)
(PNIPAM)~\cite{rodriguez2015mechanism,bischofberger2014co,schild1991cononsolvency}.

In this study, we demonstrate that the coil-globule-coil transition can be
induced using a minimalistic model where all interactions, except those between
solvent and cosolvent particles, are repulsive. Contrary to conventional theories emphasizing
preferential interactions between the polymer and one of the solvents, our
model assumes symmetric interactions between the polymer and both solvent
types. Yet, coil-to-globule transitions are observed solely by tuning
solvent-cosolvent interactions. This outcome is attributed to the emergence of
effective depletion interactions in the system. Additionally, we show that the
strength of these depletion interactions, modulated by solvent-solvent
attraction, enhances polymer collapse. Finally, we explore how differential
interactions between solvent and cosolvent contribute to coil-to-globule
transitions in systems influenced by both depletion and bridging interactions.
Furthermore, we investigate how size asymmetry between the two solvents and
temperature modulate the cononsolvency phenomenon.

\section{Model and Methods}\label{Sec-2}

We consider a coarse-grained bead-spring model for a linear homopolymer consisting of $N_p$ identical monomers ($N_p = 100$) in a simulation box($L=16.5\sigma_p$, where $\sigma_p$ is the size of monomer) of volume $V$, in the presence of a mixture of solvent ($N_s$) and cosolvent ($N_c$) particles. The total number of solvent and cosolvent particles, $N_{sc}$ (where $N_{sc} = N_s + N_c$), is set to 4000 and the total solvent number density is defined as $\phi_{sc} = \frac{N_{sc}}{V}$. For all simulations, regardless of the solvent/cosolvent size, the total density $\phi_{sc}$ is kept constant at 0.890. We varied the cosolvent fraction $x_c$ from 0.1 to 1.0. Unless otherwise stated, the monomers in the polymer, as well as the solvent and cosolvent particles, have the same size.

The adjacent monomers in the polymer are connected by harmonic springs. A pair of non-bonded particles at a distance $r$ interact through the Weeks-Chandler-Andersen (WCA) potential:
\be
V_{LJ} (r)= 4\epsilon_{ij}\left[ \left(\frac{\sigma_{ij}}{r} \right)^{12}-\left(\frac{\sigma_{ij}}{r} \right)^6 \right], r < r_c
\ee
where $i,j$ take on values of $p$, $s$, and $c$, depending on whether the particle is a monomer, solvent, or cosolvent. The interactions between intra-polymer monomers and polymer-solvent/cosolvent particles are modeled using the repulsive WCA potential, with $r_c = 1.12\sigma_p$ and $\epsilon_{pp}=\epsilon_{ss}=\epsilon_{cc}=1.0$, unless otherwise stated. The interaction between solvent and cosolvent particles is taken to be attractive, with $r_c = 3.0\sigma_{sc}$ and the strength of interaction $\epsilon_{sc}$ is varied. The monomers of the polymer are connected via harmonic springs 
\begin{equation}
V_{bond}(r)=\frac{1}{2} k (r-b)^2,
\end{equation}
where we set $b=1.12 \sigma_{p}$, $r$ the separation between the bonded particles, and $k$ is the stiffness constant. We choose $k=\frac{500 \epsilon_{pp}}{\sigma_{p}^2}$. 

The equations of motion were integrated using the MD LAMMPS software package~\cite{plimpton1995fast}, and visualization of images and trajectories was performed using the Visual Molecular Dynamics (VMD) package~\cite{HUMP96}. The time step is $\delta t = 0.001\tau$, where $\tau = \sigma_{p}\sqrt{\frac{m}{\epsilon_{pp}}}$, with $m$, $\sigma_{p}$, and $\epsilon_{pp}$ representing the units of mass, length, and energy scales, respectively. Simulations were run for $2 \times 10^7$ steps using the velocity-Verlet algorithm. All simulations were conducted under constant volume and temperature conditions using the Nose-Hoover thermostat, with temperature adjusted as required. The initial configuration was equilibrated in the NPT ensemble for $10^7$ time steps to achieve the desired density. A block average over the last $n = 900$ frames was calculated to determine the average of each parameter in each simulation, with a block length of 100, resulting in $n_b = 9$ blocks. The standard deviation of these blocks from the mean was computed and divided by $\sqrt{n_b}$ to obtain the error.

\section{\label{sec:results} Results}
\subsection{Demonstration of cononsolvency}
\begin{figure}
\includegraphics[width=\columnwidth]{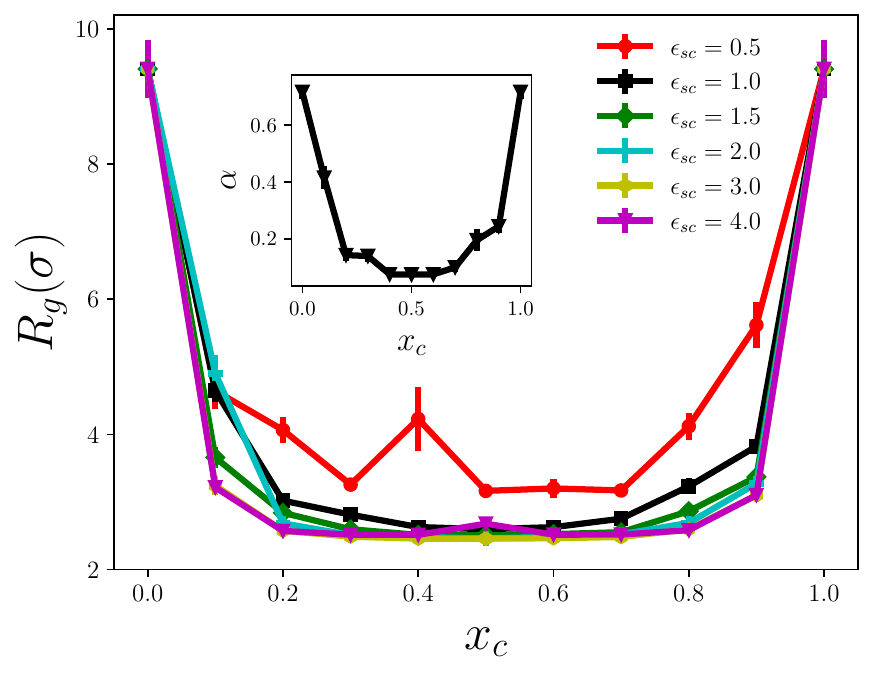}
 \caption{Radius of gyration $R_g$ as a function of cosolvent fraction
	$x_c$ for different solvent-cosolvent attractive interaction energies.
	The inset indicates the variation of the asphericity parameter with cosolvent fraction.} 
	\label{fig:Rg}
\end{figure}
We first demonstrate the effect of cononsolvency by simulating a polymer in a mixture of solvent and cosolvent, varying the cosolvent fraction $x_c$. As detailed in the Methods section, all interactions in the system are repulsive, except for the attractive interactions between solvent and cosolvent particles. When the polymer is surrounded solely by either solvent or cosolvent particles, it adopts an extended conformation, as expected. This is evident from the $R_g$ values at $x_c = 0$ and $x_c = 1$ in Figure~\ref{fig:Rg} for different solvent-cosolvent attractive strengths $\epsilon_{sc}$. Between these two extremes, as the cosolvent fraction increases, the polymer undergoes a transition from an extended to a collapsed conformation, reaching its most compact state at $x_c \sim 0.5$, which corresponds to an equal ratio $\phi_c/\phi_s=1$, where $\phi_s$ and $\phi_c$ are the volume fractions of solvent and cosolvent particles, respectively. Figure~\ref{fig:Rg}(inset) shows the variation of asphericity $\alpha$ of the polymer, along with $R_g$, as a function of cosolvent fraction $x_c$ for a solvent-cosolvent interaction strength of $\epsilon_{sc}=1.0$, further confirming this observation. For cosolvent fractions $x_c > 0.5$, the polymer exhibits a reentrant extended conformation. The results in Figure~\ref{fig:Rg} also indicate that cononsolvency is influenced by the strength of the attraction between solvent and cosolvent particles. Specifically, when the relative attraction strength between solvent and cosolvent particles is lower ($\epsilon_{sc} = 0.5$) than the strength of all other inter-particle interactions ($\epsilon_{ij} = 1.0$, where $i$ and $j$ represent polymer, solvent, and cosolvent particles), the cononsolvency effect is diminished, and the polymer does not reach a fully collapsed state.
\begin{figure}
\includegraphics[width=\columnwidth]{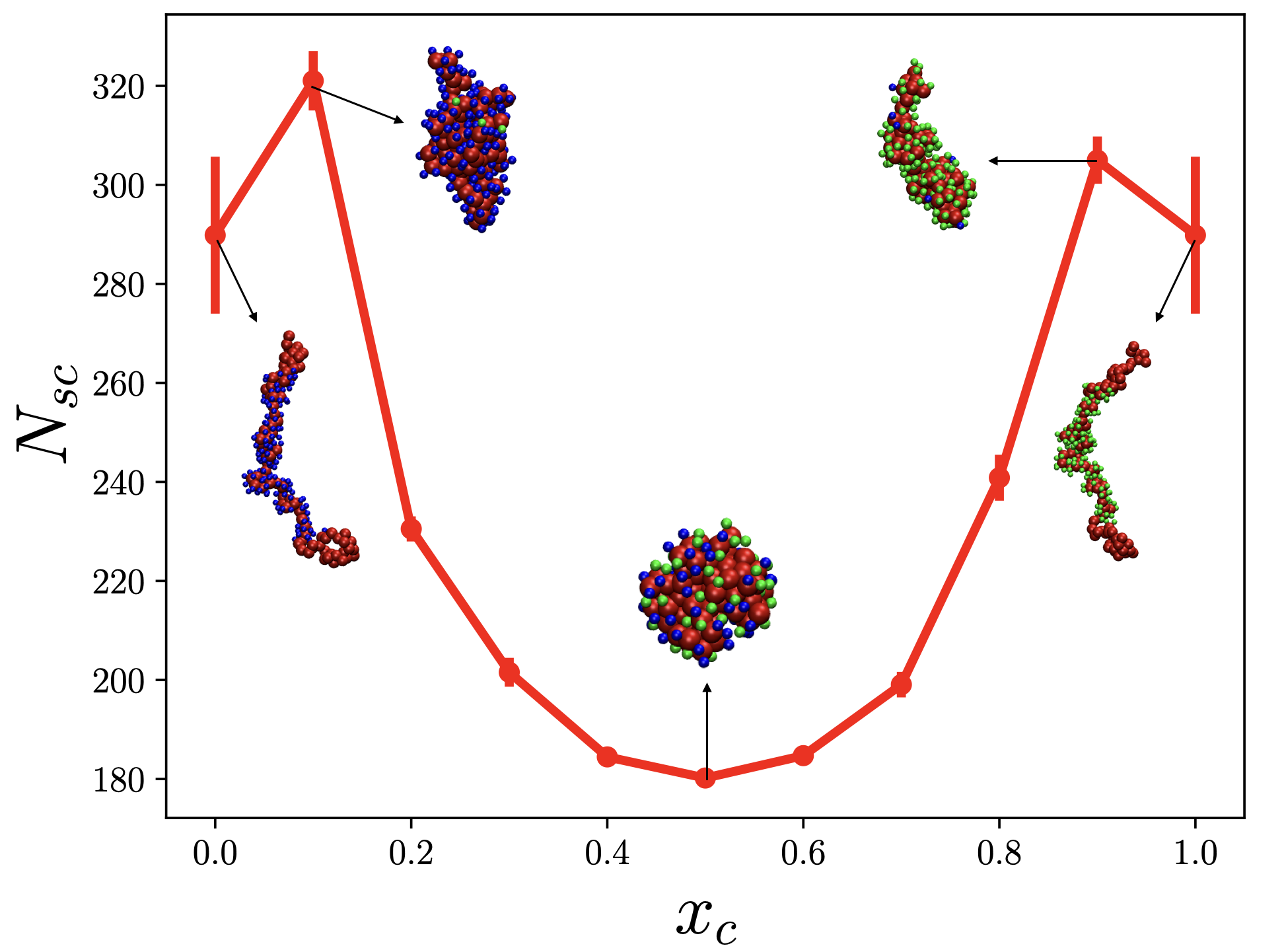}
 \caption{Total number of solvent and cosolvent particles in contact with polymer within 1.5$\sigma_s$ of chain backbone when $\epsilon_{sc}=1.0$. Snapshots of polymer (solvent and cosolvent particles colored blue and green respectively) at different cosolvent fraction $x_c$ are also shown.} 
\label{fig:noc_snapshots}
\end{figure}

We next quantify the distribution of solvent and cosolvent particles around the polymer by measuring the number of particles within a cutoff distance of $1.5\sigma_s$ from any monomer. This analysis, performed as a function of the cosolvent fraction ($x_c$), is presented in Figure~\ref{fig:noc_snapshots} for the case of $\epsilon_{sc} = 1.0$. At low cosolvent fractions ($x_c = 0.1$ or $x_c = 0.9$), an increase in the number of particles near the polymer is observed, indicating that particles are drawn toward the polymer, increasing the local density. However, as $x_c$ further increases, the attraction between solvent and cosolvent particles strengthens, leading to their depletion from the polymer surface. This depletion reaches a maximum at $x_c \sim 0.5$ ($\phi_c/\phi_s \sim 1$), coinciding with the most pronounced cononsolvency effect, as shown in Figure~\ref{fig:Rg}. Figure S1 further illustrates the individual contributions of solvent and cosolvent particles near the polymer, highlighting the crossover in their numbers at $x_c = 0.5$. These observations strongly suggest that in the present model, cononsolvency in mixed good solvents is driven by effective and enhanced depletion effects. The conformational states of the polymer, along with the surrounding solvent and cosolvent particles at various cosolvent fractions $x_c$, are also depicted in Figure~\ref{fig:noc_snapshots}. To further elucidate the interplay between solvent and cosolvent particles near the polymer as a function of $x_c$, the radial distribution function of both solvent and cosolvent particles around the polymer was computed, as shown in Figure S2. The reduction in peak heights with increasing $x_c$ further corroborates the depletion of solvent and cosolvent particles, not only near the polymer but also at greater distances, strongly supporting the role of depletion effects in good solvent mixtures within this model.

\begin{figure}
\includegraphics[width=\columnwidth]{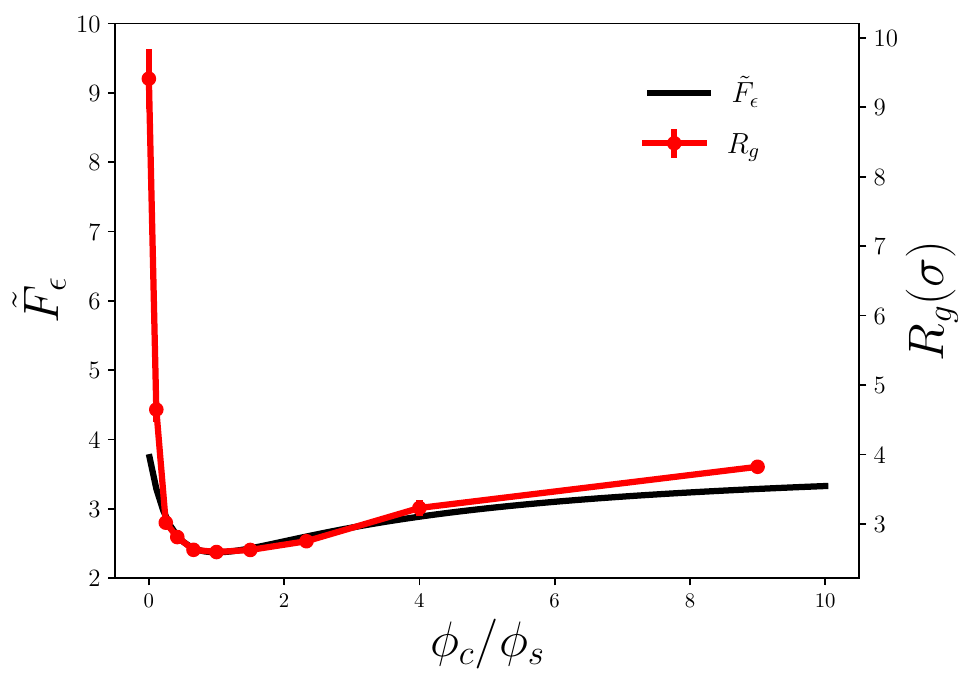}
 \caption{The varitaion of radius of gyration and energetic part of Flory-Huggins free energy with ratio of cosolvent and solvent fraction , $\frac{\phi_c}{\phi_s}$. $\tilde{F}_{\epsilon}$  is shifted by 7.0} 
\label{fig:rgw_ratio}
\end{figure}

The theoretical explanation for this observation can be drawn from the Flory-Huggins lattice theory, as discussed by Dudowicz et al.~\cite{dudowicz2015communication}. Within this framework, the total Helmholtz free energy density, $\tilde{F}_{\text{total}}$, consists of two main contributions: the entropic term, arising from the combinatorial free energy density, $\tilde{F}_{\text{comb}}$, and the enthalpic term, $\tilde{F}_{\epsilon}$, which accounts for interaction energies. The enthalpic free energy density $\tilde{F}_{\epsilon}$ for homopolymer solutions in a mixture of solvent ($s$) and cosolvent ($c$) on a lattice can be expressed as:
\begin{equation}
\begin{array}{ll}
\tilde{F}_{\epsilon}= &-\frac{h}{2k_BT}[\epsilon_{pp}\phi_{p}^2+\epsilon_{ss}\phi_{s}^2+\epsilon_{cc}\phi_{c}^2
\\
& +2\epsilon_{ps}\phi_{p}\phi_{s}+2\epsilon_{pc}\phi_{p}\phi_{c}+2\epsilon_{cs}\phi_{c}\phi_{s}]
\end{array}
\end{equation}
where $\phi_p$, $\phi_c$, and $\phi_s$ represent the volume fractions of the polymer, cosolvent, and solvent, respectively. Here, $h$ is the coordination number of the lattice, $T$ is the temperature, and $\epsilon_{\alpha\beta}$ represents the interaction energy between neighboring species $\alpha$ and $\beta$. By plotting $\tilde{F}_{\epsilon}$ and $\tilde{F}_{\text{total}}$ as functions of the cosolvent-to-solvent ratio, $\phi_c/\phi_s$, while keeping $\phi_p$ constant and assuming attractive solvent-cosolvent interactions, a minimum in the free energy emerges when the cosolvent and solvent fractions are equal, $\phi_c/\phi_s = 1$, as shown in Figure S3 (see SI). The location of the minimum in the variation of the enthalpic free energy $\tilde{F}_{\epsilon}$ (shifted vertically by a constant value of 7.0) is in very good agreement with the minimum in $R_g$ obtained from our simulations, as shown in Figure~\ref{fig:rgw_ratio} for $\phi_c/\phi_s \sim 1$, corresponding to a cosolvent fraction of $x_c \sim 0.5$. Calculations carried out by Zhang et al.~\cite{zhang2020unified}, using the random phase approximation, also suggest that solvent quality is symmetric around $\phi_c/\phi_s \sim 1$ for an equal-quality binary mixture, consistent with the results discussed earlier. These findings suggest that the cononsolvency behavior, in the present model, are driven by enthalpic contributions, in contrast to pure depletion interactions, which are entropic in origin.

\subsection{Role of solvent-cosolvent particle sizes on cononsolvency}
\begin{figure}
\includegraphics[width=\columnwidth]{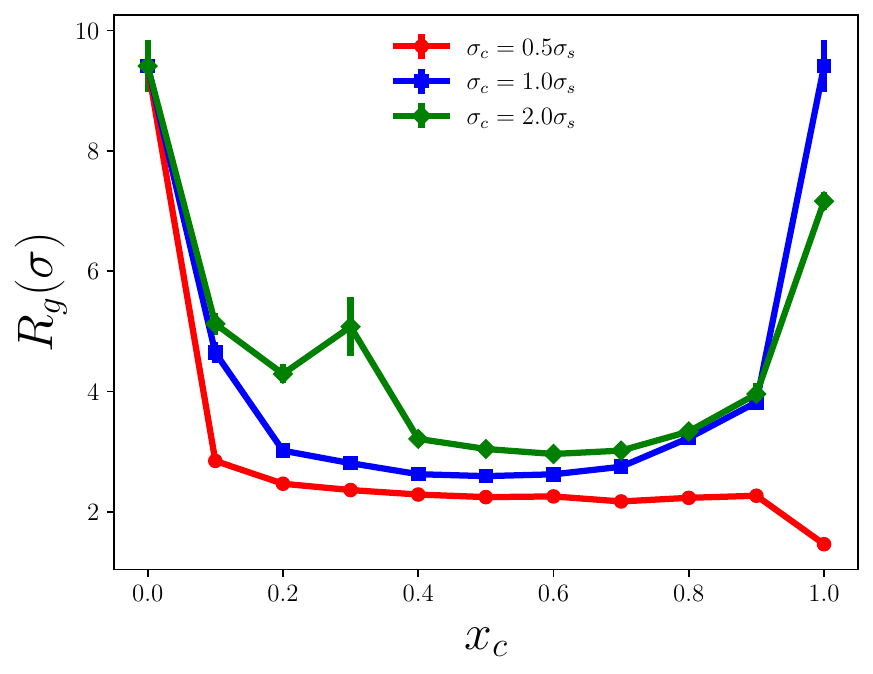}
 \caption{Radius of gyration $R_g$ as a function of cosolvent fraction $x_c$ for different cosolvent particle sizes. } 
\label{fig:Rg_diff_sigma}
\end{figure}
\begin{figure}
\includegraphics[width=\columnwidth]{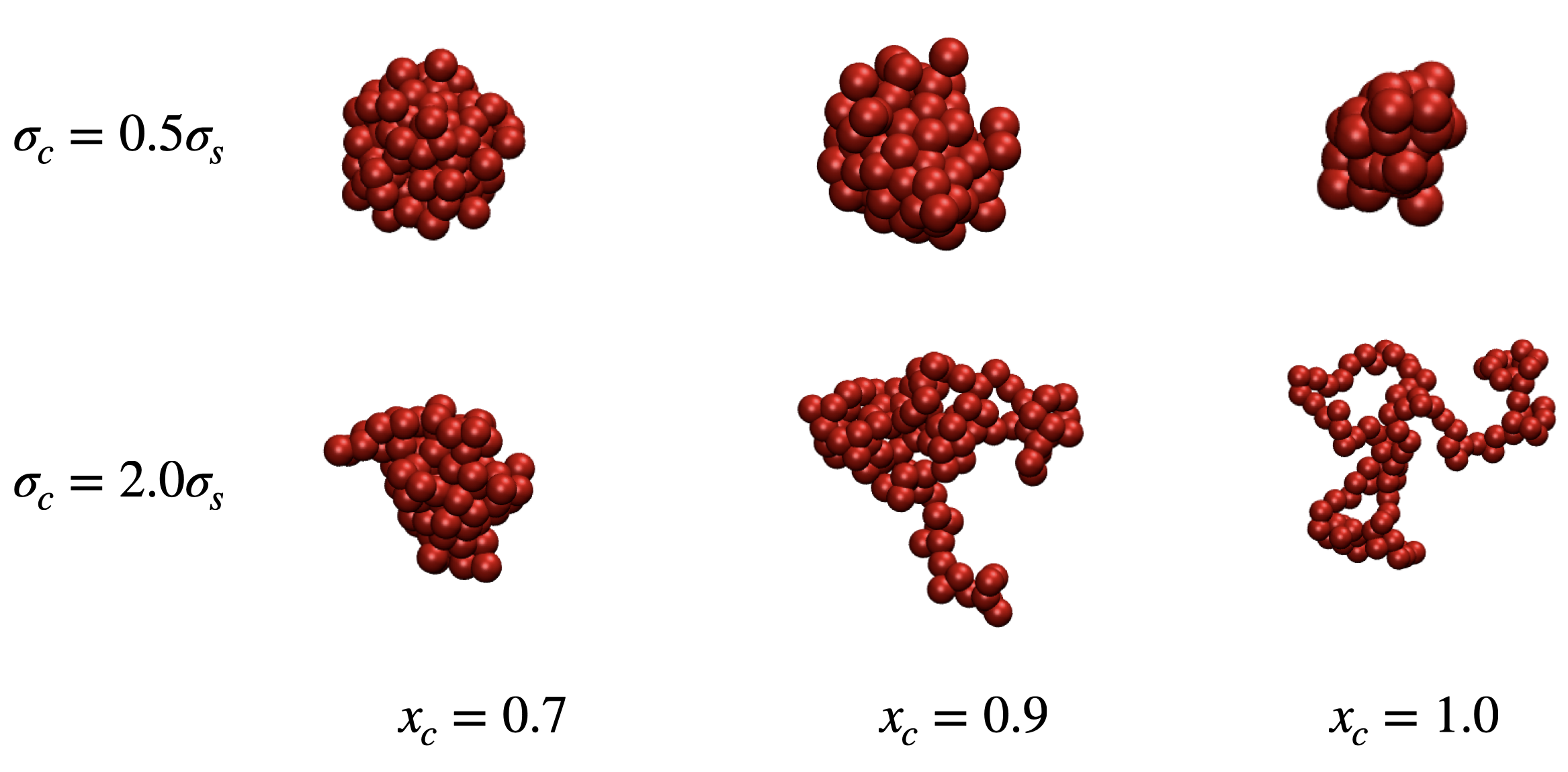}
 \caption{Snapshots of polymer at different cosolvent fraction $x_c$ for cosolvent size $\sigma_c=0.5\sigma_s$ and $\sigma_c=2.0\sigma_s$.} 
\label{fig:snapshots_sA_half}
\end{figure}

The role of solvent and cosolvent sizes on polymer conformation is critical, as size asymmetry can significantly influence polymer behavior~\cite{bharadwaj2020cosolvent,yong2018cononsolvency,lekkerkerker1992phase,louis2002effective,kim2015polymer,jeon2016effects,yodh2001entropically}. To investigate the effects of size asymmetry between solvent and cosolvent particles on the cononsolvency phenomenon, we consider three cosolvent sizes: $\sigma_c = 0.50$, $1.00$, and $2.00$, while keeping the solvent particle size, $\sigma_s$, and monomer size, $\sigma_p$, constant and same. As shown in Figure~\ref{fig:Rg_diff_sigma}, cosolvent sizes $\sigma_c = 1.0$ and $2.0$ exhibit cononsolvency behavior in the presence of solvent-cosolvent attraction, whereas $\sigma_c = 0.5$ results in the polymer remaining in a collapsed state at all cosolvent fractions. The volume density for $\sigma_c = 0.5$ exceeds the depletion threshold —leading to the polymer maintaining its collapsed conformation as soon as cosolvent particles are introduced. Snapshots at higher cosolvent fractions ($x_c>0.6$) for $\sigma_c = 0.50$ and $\sigma_c = 2.00$ in Figure~\ref{fig:snapshots_sA_half} illustrate this behavior: the polymer adopts a reentrant extended structure for large cosolvent particles ($\sigma_c = 2.00$), while maintaining a very compact structure for small cosolvent particles ($\sigma_c = 0.50$).

Importantly, the size asymmetry with the monomer is also a key factor, not just the size difference between solvent and cosolvent particles. As the cosolvent fraction $x_c$ increases, the spatial distribution of solvent and cosolvent particles near the polymer surface varies, as illustrated by the radial distribution functions $g_{pc}$ and $g_{ps}$ for two different cosolvent particle sizes in Figure S4. For the $\sigma_c = 0.5\sigma_s$ case, increasing the cosolvent fraction leads to greater accumulation of cosolvent particles near the polymer, reaching a maximum at $x_c = 1$, driving the polymer into a highly compact state. In contrast, for $\sigma_c = 2.0\sigma_s$, no significant accumulation occurs near the polymer surface, even as $x_c$ increases, allowing the polymer to remain extended. This can also be seen in the crossover of the number of solvent and cosolvent particles near the polymer as shown in Figure S5. For $\sigma_c = 0.5\sigma_s$, the crossover occurs at much smaller cosolvent fraction compared to the $\sigma_c = 2.0\sigma_s$ case. This behavior highlights how size asymmetry between solvent and cosolvent enhances depletion interactions when larger particles (solvent) cannot fit into regions between a polymer and smaller cosolvent particles. This creates an imbalance in osmotic pressure that exerts an attractive force between polymer segments, promoting collapse when the cosolvent size is smaller than the monomer. However, when the solvent and cosolvent particles are similar in size or larger than the monomer, size asymmetry does not lead to increased compaction, and the earlier observed reentrant extension occurs. This behavior reflects the complex interplay of polymer-crowder interactions, heavily influenced by both the size and density of crowders. For instance, Kim et al.\cite{kim2015polymer} examined the effect of crowder size on polymer conformation in cylindrical confinement and found that polymer chains collapse as the volume fraction of crowders increases, particularly when the crowder size is less than 0.6 times the monomer size. This aligns with classical polymer-crowder interaction theory, which posits that the free energy gain due to crowder exclusion is proportional to the polymer’s surface area exposed to the solvent. Similarly, Jeon and colleagues\cite{jeon2016effects} demonstrated that, for crowder sizes smaller than the monomer, polymer compaction is dependent on both the crowder size and density, whereas for larger crowders, compaction is primarily influenced by crowder density. This phenomenon is consistent with generalized depletion theory, which predicts that smaller crowders are more effective at inducing polymer collapse due to their ability to penetrate closer to the polymer surface.

\subsection{Temperature and cononsolvency}
  \begin{figure}
\includegraphics[width=\columnwidth]{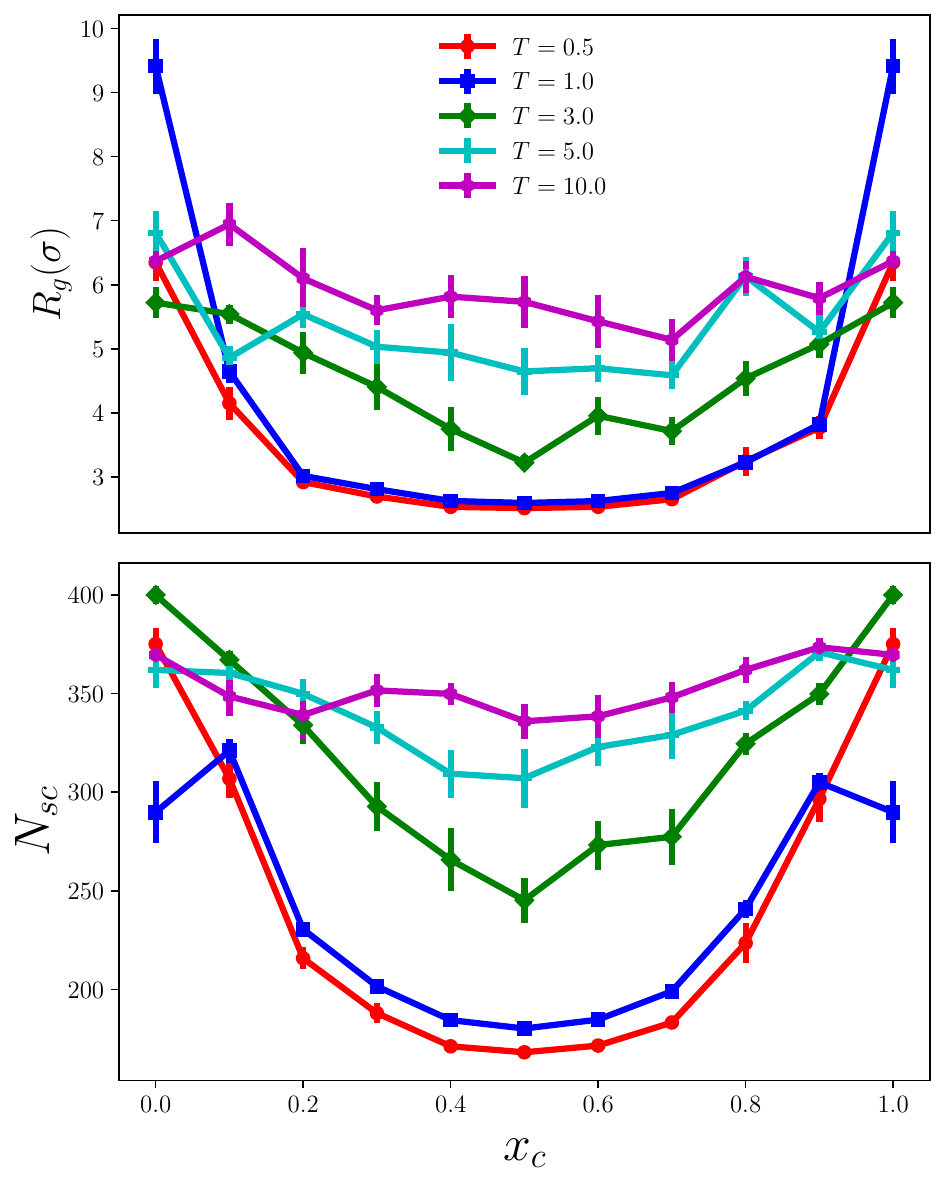}
 \caption{(a) Radius of gyration $R_g$ as a function of cosolvent fraction $x_c$ different temperature.(b) Number of solvents/cosolvents in contact with polymer within $1.5\sigma_s$ of polymer backbone. } 
\label{fig:Rg_noc_diff_temp}
\end{figure}
\begin{figure}
\includegraphics[width=\columnwidth]{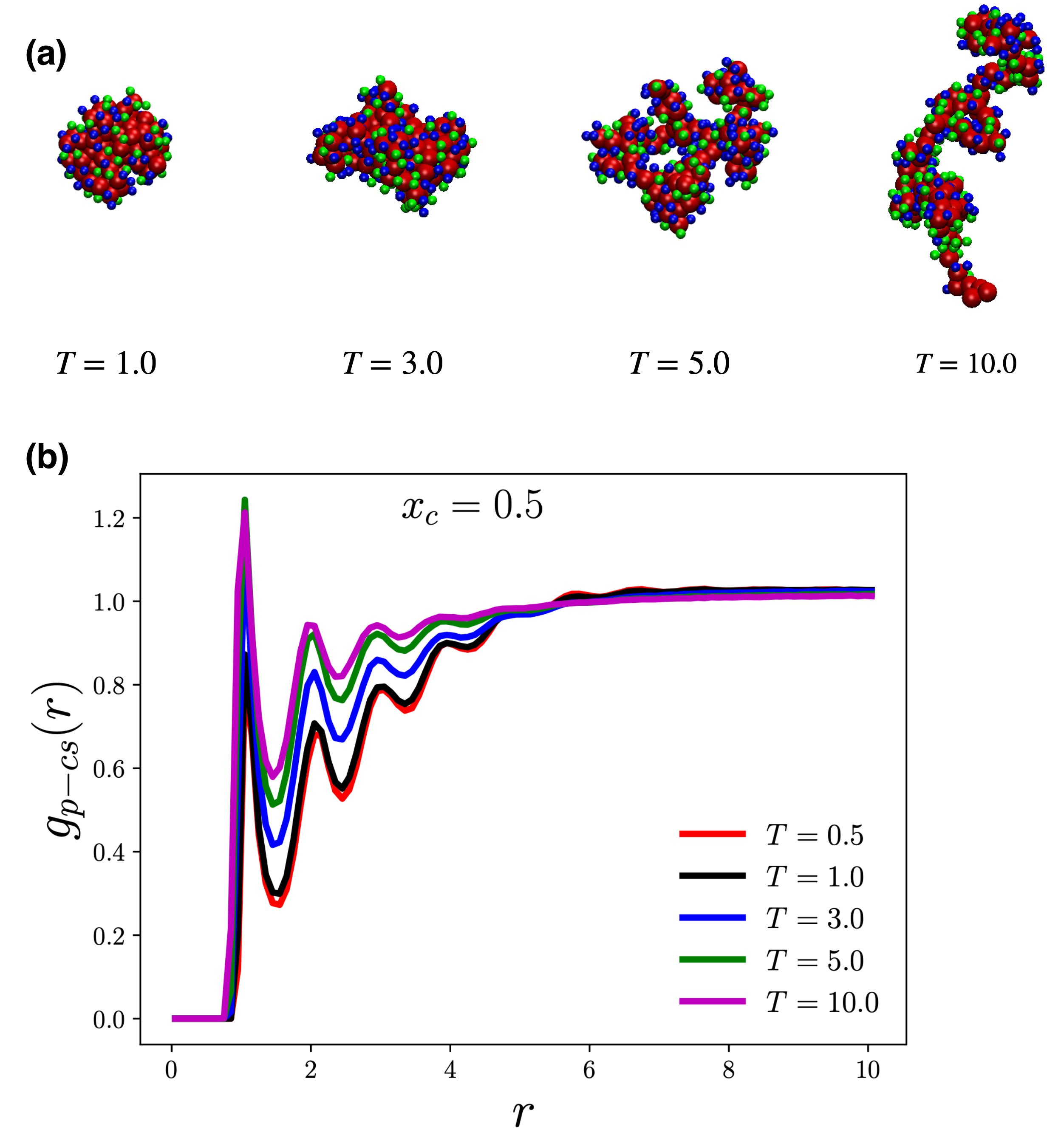}
 \caption{(a) The snapshots of polymer(solvent and cosolvent particles colored blue and green respectively) at different temperatures for cosolvent fraction $x_c=0.5$ (b) Radial distribution function between polymer and solvent/cosolvent for cosolvent fraction $x_c=0.5$ } 
\label{fig:gofr_diff_temp}
\end{figure}

Temperature plays a crucial role in determining the mixing behavior of solutions. In this study, we investigate the effect of temperature on the cononsolvency phenomenon driven by solvent-cosolvent attraction. The variation in the radius of gyration ($R_g$) with the cosolvent fraction ($x_c$) at different temperatures is presented in Figure~\ref{fig:Rg_noc_diff_temp}(a). As temperature increases, we observe a diminishment in the cononsolvency effect. A similar trend is observed in the number of solvent and cosolvent particles in the vicinity of the polymer chain (Figure~\ref{fig:Rg_noc_diff_temp}(b)), which reconfirms that, in the current model, cononsolvency is driven primarily by enhanced depletion effects arising from solvent-cosolvent attraction. At higher temperatures, the increased thermal energy results in greater entropy, causing particles to become more uniformly distributed around the polymer as opposed to lower temperatures, where both solvent and cosolvent particles are more excluded from the polymer surface. This leads to an extended polymer conformation at elevated temperatures, as seen from the conformations corresponding to $x_c = 0.5$, where cononsolvency is most pronounced (Figure~\ref{fig:gofr_diff_temp}(a)). Snapshots and radial distribution functions for solvent and cosolvent particles near the polymer surface (Figure~\ref{fig:gofr_diff_temp}(b)) also illustrate that, at lower temperatures, the particles are excluded from the polymer surface, while at higher temperatures, they decorate the polymer, leading to a more extended state.

\begin{figure}
\includegraphics[width=\columnwidth]{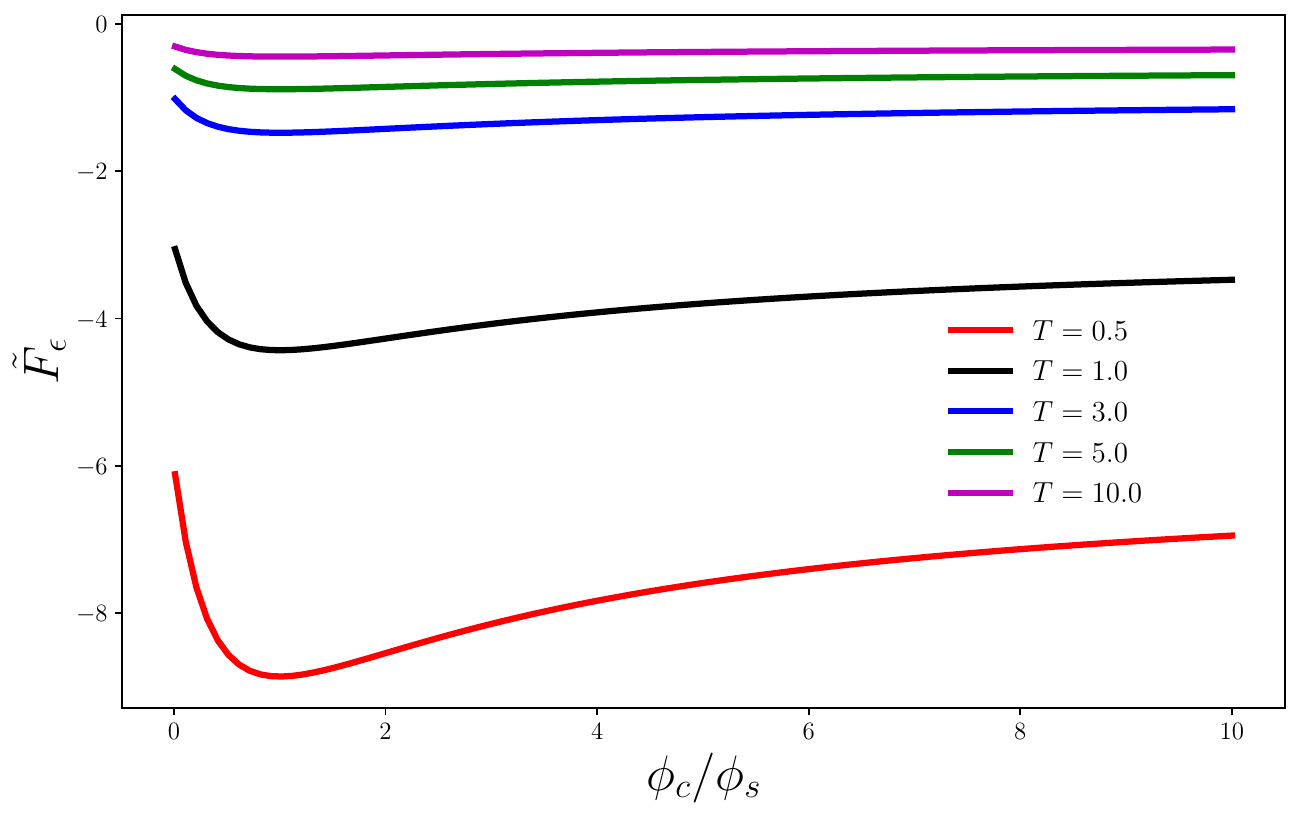}
 \caption{The effect of temperature on interaction free energy from Flory-Huggins theory with ternary mixture. The plot shows as temperature increases, the contribution of inetraction free energy decreases and consistent with simulation results } 
\label{fig:fhenergy_withT}
\end{figure}

To further investigate the effect of temperature on the cononsolvency behavior, we examine the enthalpic free energy contribution in the Flory-Huggins theory ($\tilde{F}_{\epsilon}$), shown in Figure~\ref{fig:fhenergy_withT}. The plot demonstrates that as temperature increases, the enthalpic contribution to the free energy decreases, leading to the disappearance of the minima in the free energy landscape, which is consistent with our simulation results. This suggests that the depletion effect becomes less pronounced at higher temperatures, and the cononsolvency phenomenon fades. The reduction in depletion effects with increasing temperature has also been observed experimentally by Feng et al.\citep{feng2015re}. Their study of colloid-polymer mixtures (Dextran and Polystyrene) showed that depletion is reduced at higher temperatures, particularly in systems where the interaction potential includes a soft component. Harries et al.\cite{sukenik2013balance,sapir2014origin,sapir2015depletion,sapir2015macromolecular,sapir2016macromolecular} used Flory-Huggins energy expressions to explain how parameters such as cosolvent size, particle interactions, and mixing behavior can cause depletion to be either enthalpic or entropic. In certain cases, depletion may be entropically disfavored but enthalpically favorable.

Experimental studies of protective osmolytes, such as trehalose and sorbitol, have shown that depletion can be enthalpically favorable but entropically unfavorable~\cite{politi2010enthalpically,sukenik2013diversity,sukenik2013balance}. Harries and colleagues developed an entropy-energy plot to elucidate the balance between entropic and energetic contributions to the stabilization free energy of the system. By incorporating non-zero entropic components in the free energy mixing parameters, they were able to access the enthalpic stabilization sector of the phase diagram. Even when solvent-cosolvent interactions are entropically disfavored on a microscopic scale, the energetic gain may still be sufficient to drive depletion. Our simple polymer model effectively captures the temperature dependence of cononsolvency and shows that the reduction in depletion effects with increasing temperature causes the disappearance of cononsolvency, particularly in the absence of preferential attraction between the polymer and cosolvent.

\subsection{Preferential attraction and cononsolvency}
\begin{figure}
\includegraphics[width=\columnwidth]{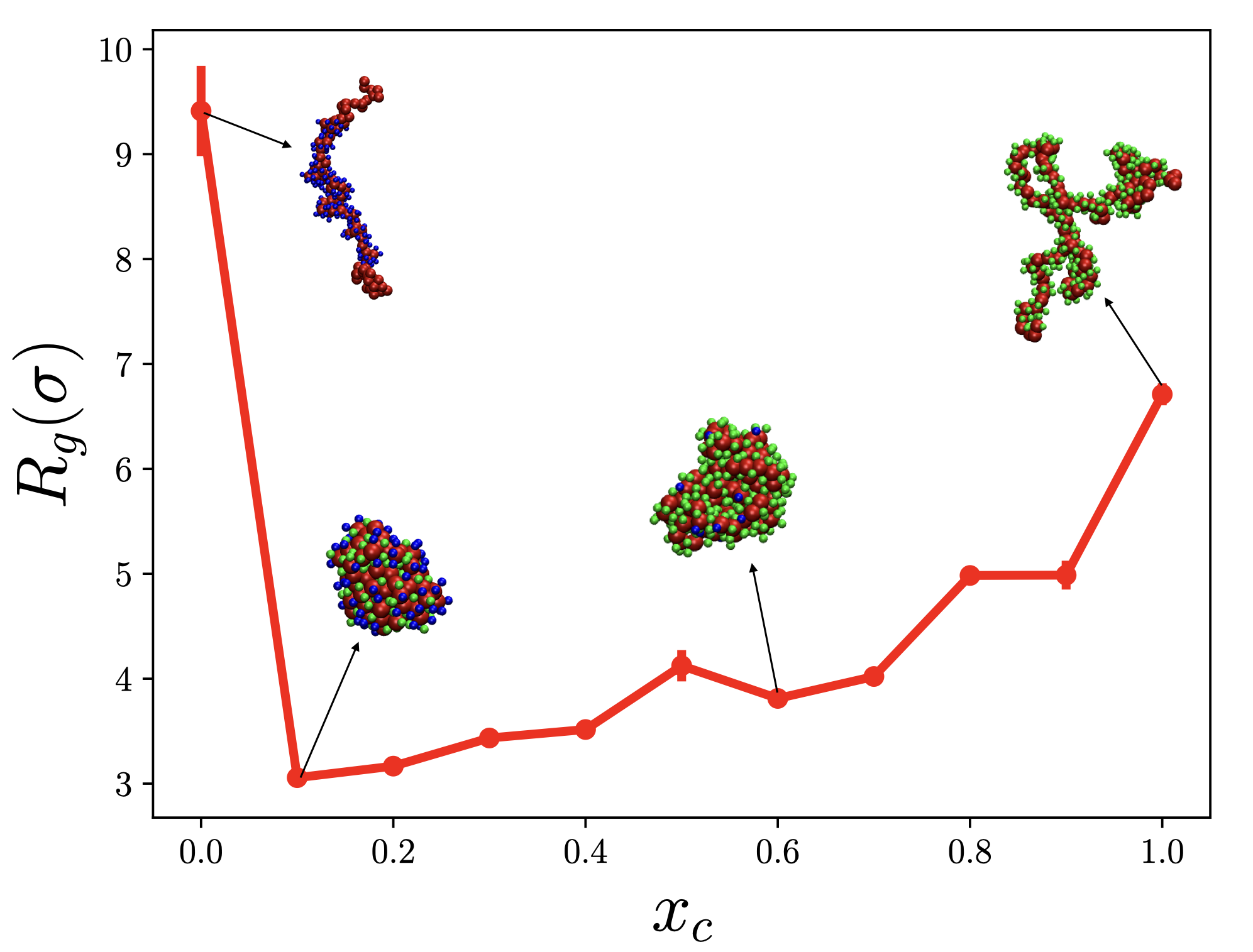}
 \caption{Radius of gyration $R_g$ as a function of cosolvent fraction $x_c$ for preferential attraction.Attractive strength between monomer-coslovent is $\epsilon_{pc}=4.0$ .Snapshots of polymer (solvent and cosolvent particles colored blue and green respectively) at different cosolvent fraction $x_c$ are also shown.} 
\label{fig:Rg_preferential_att}
\end{figure}

In this section, we introduce preferential attraction between polymer and cosolvent, in addition to the existing attractive interactions between solvent and cosolvent particles, and explore its effects on cononsolvency within the present model. The preferential attraction of cosolvent particles to polymer segments can lead to effective bridging interactions. This occurs when cosolvent particles simultaneously interact with multiple polymer segments, effectively pulling them closer and promoting collapse, as seen in previous studies~\cite{garg2023conformational,heyda2013rationalizing,tripathi2023conformational,antypov2008computer}. We model the preferential polymer-cosolvent attraction using the Lennard-Jones potential with an interaction strength $\epsilon_{pc} = 4.0$, based on previous work where bridging interactions were shown to be effective~\cite{garg2023conformational}. The results of the polymer conformations, expressed through the radius of gyration ($R_g$) as a function of cosolvent fraction $x_c$, along with corresponding snapshots, are shown in Figure~\ref{fig:Rg_preferential_att}. As soon as attractive cosolvent particles are introduced, the polymer undergoes a coil-to-globule transition, driven by bridging interactions—an effective polymer-polymer attraction mediated by cosolvent particles. However, as the fraction of attractive cosolvent particles increases, the polymer-cosolvent interaction becomes increasingly preferable, leading to a globule-to-coil transition. This behavior aligns with the results of Mukherji et al.~\cite{mukherji2014polymer}, who reported a similar smooth transition, albeit using a different model.
While the preferential attraction of cosolvent to the polymer also induces cononsolvency, the underlying mechanism here differs significantly from the depletion-driven mechanism discussed in previous sections. In those cases, chain collapse was driven by enhanced depletion effects. In contrast, in the current scenario, cononsolvency is driven by bridging interactions~\cite{mukherji2014polymer}. In our earlier work~\cite{garg2023conformational}, where we modeled the solvent implicitly, the polymer exhibited a coil-to-globule transition as the attraction between polymer and cosolvent (crowders) increased. The reentrant behavior observed as the cosolvent fraction increases can be attributed to the fact that, as more cosolvents decorate the polymer chain, the polymer begins to favor an extended state.  The location of the minimum in solvent quality,In the case of preferential attraction as suggested by Zhang et al.~\cite{zhang2020unified}, is given by:
\begin{equation}
x_c^*=\frac{1}{2}-\frac{2}{c\Delta\chi}
\end{equation}
where $c = 1 - \phi_p$ and $\Delta \chi = \chi_{ps} - \chi_{pc}$, where $\chi_{ij}$ is Flory interaction parameter between component $i$ and $j$~\cite{10.1093/oso/9780198520597.001.0001}. For the preferential attraction case, $\Delta \chi$ is a non-zero positive value, leading to $x_c^* < 0.5$, which agrees with our simulation results. This finding is consistent with earlier studies where the collapse transition of a model polymer occurred at a lower cosolvent fraction for ethanol due to higher preferential adsorption of ethanol on the collapsed polymer globule~\cite{bharadwaj2020cosolvent,walter2012molecular}. Additionally, stronger polymer-cosolvent attractions result in more extended polymer states beyond $x_c = 0.2$ in our work, explaining why poly(diethylacrylamide) (PDEA) does not exhibit cononsolvency in PDEA-methanol-water solutions~\cite{hofmann2013cononsolvency,liu2015phase,maeda2002change,scherzinger2010cononsolvency}.

\section{\label{sec:summary} Discussion and Conclusion}

We present a comprehensive study on the cononsolvency effect in linear polymers dissolved in a mixture of good solvents. Our simulation results demonstrate that solvent-cosolvent attraction alone is sufficient to induce the collapse of polymer chains in a mixture of two good solvents. Furthermore, our findings suggest that cononsolvency is a generic phenomenon, not restricted to specific chemical systems. When the polymer is dissolved in a solution containing only solvents, the chain adopts an extended conformation, as expected under good solvent conditions. Introducing cosolvents ($c$) with Weeks-Chandler-Andersen (WCA) repulsive interactions with solvents does not alter this extended conformation. However, when solvent-cosolvent interactions become attractive, the polymer chain undergoes collapse as the cosolvent fraction increases, displaying cononsolvency behavior. The polymer reaches a fully compact state when the fractions of solvent and cosolvent are equal. This cononsolvency behavior is driven by the effective, enhanced depletion effect observed in our model. As the cosolvent fraction increases, both solvent and cosolvent particles are excluded from the polymer surface, leading to chain collapse. The attraction between solvent and cosolvent particles amplifies this depletion effect. In contrast, in other systems, solvent-cosolvent repulsion leads to a depleted depletion effect, resulting in cosolvency rather than cononsolvency~\cite{mukherji2017depleted}.

The Flory-Huggins mean-field theory has been employed to explain cononsolvency behavior driven by solvent-cosolvent attraction~\cite{dudowicz2015communication}. Previous studies have derived expressions for the effective Flory-Huggins interaction parameter, showing that attractive interactions between solvent and cosolvent promote immiscibility~\cite{kambour1983phase,ten1983phase,dudowicz1995modification}. According to this theory, the minima of the free energy occur when the fractions of solvent and cosolvent are equal ($x_s = x_c = 0.5$), corresponding to a relative solvent composition $\phi_c/\phi_s = 1$. This is consistent with our simulation results, where the variation in the polymer’s radius of gyration follows the same trend as the energetic part of the free energy derived from Flory-Huggins theory. Plots of free energy as a function of relative solvent composition $\phi_c/\phi_s$ suggest that the energetic term predominantly drives this behavior. Similarly, calculations by Zhang et al.~\cite{zhang2020unified} using the random phase approximation indicate that solvent quality is symmetric around $\phi_c/\phi_s = 1$ for an equal-quality binary mixture, as predicted by Flory-Huggins theory and observed in our simulations. Unlike earlier simulations that used more complex models with specific interactions between polymer and solvent/cosolvent particles to demonstrate cononsolvency~\cite{mukherji2014polymer,dalgicdir2017computational,bharadwaj2021interplay,tavagnacco2020molecular,perez2019p} , our results show that even a simple polymer model with minimal interactions can accurately match theoretical predictions and exhibit cononsolvency. In our system, cononsolvency is achieved without specialized interactions between the polymer and solvent particles, implying that modifying the effective attractions between solvent and cosolvent particles alone is sufficient to induce the phenomenon.

The size of solvent and cosolvent particles plays a crucial role in determining polymer conformation, particularly in driving cononsolvency behavior through solvent-cosolvent attraction. In our simulations, we find that when the cosolvent size is smaller than the polymer monomer size ($\sigma_c = 0.5\sigma_s$), the polymer remains collapsed for most cosolvent fractions, as the volume density exceeds the threshold necessary for inducing collapse. This behavior is consistent with depletion theory, which suggests that smaller particles create stronger depletion forces, leading to more effective polymer collapse\cite{jeon2016effects,kim2015polymer,kang2015effects,sharp2015analysis}. For larger cosolvent sizes ($\sigma_c = 1.0$ and $2.0$), the polymer exhibits typical cononsolvency behavior—collapsing at intermediate cosolvent fractions and re-expanding at higher fractions. While the role of particle size in depletion effects has been extensively studied, such as by Kang et al.\cite{kang2015effects} and Sharp et al.\cite{sharp2015analysis}, our results highlight that solvent-cosolvent attraction alone is sufficient to drive cononsolvency. In systems where the solvent and cosolvent particle sizes are comparable to the polymer monomer size, we observe that the polymer's conformation is highly sensitive to both size and the relative volume fraction of cosolvent. Jeon et al.\cite{jeon2016effects} demonstrated that for crowders smaller than the polymer monomers, compaction is driven by both size and density, while larger crowders primarily depend on their density. The work of Kim et al.\cite{kim2015polymer} and others emphasizes that smaller particles generate stronger depletion effects, which our study corroborates. Our results align with these findings, as smaller cosolvent particles induce polymer collapse at lower volume densities, while larger particles promote cononsolvency. This size-dependent behavior is critical for understanding cononsolvency in mixed solvent environments, where the interplay between solvent-cosolvent interactions and crowding effects becomes complex. We extend previous findings by showing that cononsolvency can be driven solely by adjusting solvent-cosolvent attraction and particle size, without requiring complex interaction models.

Traditionally, cononsolvency has been linked to smart polymers exhibiting lower critical solution temperature (LCST) behavior, such as PNIPAm, PVCL, or PAPOMe~\cite{tanaka2008temperature,hiroki2001volume}. This connection suggests that cononsolvency could be temperature-sensitive. Our simulations confirm this, showing that as temperature increases, the cononsolvency effect weakens. This reduction in cononsolvency corresponds to a more uniform distribution of solvent and cosolvent particles around the polymer chain, indicating that the depletion effect diminishes with rising temperature. This observation aligns with experimental findings from Feng et al.~\cite{feng2015re}, who reported a similar decrease in depletion-induced polymer collapse in colloid-polymer mixtures at higher temperatures. Harries et al.~\cite{sukenik2013balance, sapir2014origin} used Flory-Huggins theory to show how depletion effects can be driven by both enthalpic and entropic contributions, depending on cosolvent size and interaction parameters. Certain depletants, such as trehalose and sorbitol, have been found to induce depletion primarily through enthalpic contributions. In our study, we observe a similar trend. As temperature increases, the interaction free energy term ($\tilde{F}_{\epsilon}$) from Flory-Huggins theory decreases, indicating that the depletion effect is largely enthalpic and weakens with rising temperature. This temperature sensitivity underscores the role of enthalpic contributions in driving cononsolvency, further corroborating the temperature dependence observed in previous studies.

Previous studies on cononsolvency have primarily focused on systems where the cosolvent preferentially interacts with the polymer chain~\cite{mukherji2014polymer}. In such cases, cononsolvency is typically attributed to mechanisms like bridging interactions, where cosolvent particles act as connectors between different segments of the polymer, inducing collapse~\cite{heyda2013rationalizing,garg2023conformational,tripathi2023conformational}. Another explanation involves the surfactant-like behavior of cosolvents, where the cosolvent reduces the free energy penalty associated with forming a repulsive polymer-solvent interface, driving polymer collapse at lower cosolvent fractions~\cite{bharadwaj2020cosolvent}. In contrast, our study demonstrates that cononsolvency can occur even in the absence of preferential interactions between the polymer and solvent/cosolvent. In our minimalistic model, cononsolvency arises solely from solvent-cosolvent attraction. However, introducing preferential interactions is an easier, more intuitive way to explore this phenomenon. When preferential attraction between the cosolvent and polymer is added, we observe cononsolvency at lower cosolvent fractions ($x_c < 0.5$), which aligns with previous work~\cite{zhang2020unified}. Our results are consistent with predictions from the random phase approximation, which suggests that the minima in polymer conformation for systems with preferential interactions occur in the range of $0.1 \leq x_c < 0.5$, as we also observe in our simulations. Thus, while our model without preferential interactions demonstrates cononsolvency through solvent-cosolvent attraction alone, introducing these interactions leads to a different mechanism of cononsolvency that occurs at lower cosolvent fractions. This finding provides a direct comparison to prior studies, emphasizing that preferential interactions, while not necessary to induce cononsolvency, shift the cosolvent fraction at which polymer collapse occurs.

Cononsolvency challenges the expectation that a polymer, which remains extended in a good solvent, should behave similarly in a mixture of good solvents. Through a simple polymer model, we demonstrate that solvent-cosolvent attraction alone is sufficient to induce cononsolvency, without the need for complex interactions. While our findings contribute to the understanding of this phenomenon, developing a unified framework to predict cononsolvency across various systems and different polymer-solvent-cosolvent interactions, as well as refining this minimal model to explore other specific aspects, remains to be further explored.

\section*{Acknowledgment}
 We thank the HPC facility at the Institute of Mathematical Sciences for providing computing time. SV is grateful for the insightful and stimulating initial discussions with R. Rajesh, which critically contributed to the development of this work.

\bibliography{cononsolvency}

\end{document}